\documentclass[aps,prd,noeprint,twocolumn,showpacs,amsmath,amssymb]{revtex4-2}
\usepackage{amsmath}
\usepackage{graphicx}
\usepackage{subfigure}
\usepackage{epstopdf}
\usepackage{color}
\usepackage{multirow}
\usepackage{setspace}
\usepackage{overpic}
\usepackage{amssymb}
\usepackage[bookmarksnumbered, pdfstartview=FitH,colorlinks,urlcolor=blue, citecolor=blue,linkcolor=blue]{hyperref}
\usepackage{lineno}
\usepackage{bm}
\usepackage{rotating}
\usepackage[utf8]{inputenc}
\let\oldequation\equation
\let\oldendequation\endequation

\renewenvironment{equation}
  {\linenomathNonumbers\oldequation}
  {\oldendequation\endlinenomath}

\begin{document}
%\linenumbers

\title{\bf \boldmath
Amplitude Analysis of Singly Cabibbo-Suppressed Decay \texorpdfstring{$\Lambda^{+}_{c}\to p K^{+} K^{-}$}{Lc2pKK}}

\newcommand{\BESIIIorcid}[1]{\href{https://orcid.org/#1}{\hspace*{0.1em}\raisebox{-0.45ex}{\includegraphics[width=1em]{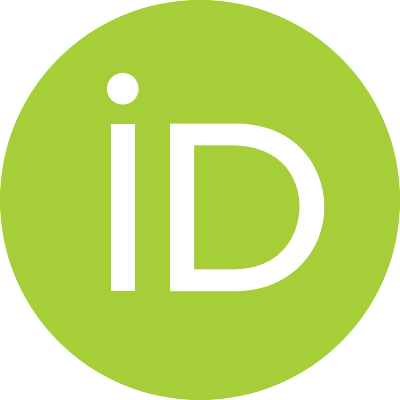}}}}

\author{
\begin{small}
\begin{center}
%% Saved at => 2025-07-28
M.~Ablikim$^{1}$\BESIIIorcid{0000-0002-3935-619X},
M.~N.~Achasov$^{4,c}$\BESIIIorcid{0000-0002-9400-8622},
P.~Adlarson$^{82}$\BESIIIorcid{0000-0001-6280-3851},
X.~C.~Ai$^{87}$\BESIIIorcid{0000-0003-3856-2415},
R.~Aliberti$^{39}$\BESIIIorcid{0000-0003-3500-4012},
A.~Amoroso$^{81A,81C}$\BESIIIorcid{0000-0002-3095-8610},
Q.~An$^{78,64,\dagger}$,
Y.~Bai$^{62}$\BESIIIorcid{0000-0001-6593-5665},
O.~Bakina$^{40}$\BESIIIorcid{0009-0005-0719-7461},
Y.~Ban$^{50,h}$\BESIIIorcid{0000-0002-1912-0374}
% ...
\\
\vspace{0.2cm}
(BESIII Collaboration)\\
\vspace{0.2cm} {\it
$^{1}$ Institute of High Energy Physics, Beijing 100049, People's Republic of China\\
% ...
$^{4}$ Budker Institute of Nuclear Physics SB RAS (BINP), Novosibirsk 630090, Russia\\
% ...
$^{39}$ Johannes Gutenberg University of Mainz, Johann-Joachim-Becher-Weg 45, D-55099 Mainz, Germany\\
$^{40}$ Joint Institute for Nuclear Research, 141980 Dubna, Moscow region, Russia\\
% ...
$^{50}$ Peking University, Beijing 100871, People's Republic of China\\
% ...
$^{62}$ Southeast University, Nanjing 211100, People's Republic of China\\
% ...
$^{64}$ State Key Laboratory of Particle Detection and Electronics, Beijing 100049, Hefei 230026, People's Republic of China\\
% ...
$^{78}$ University of Science and Technology of China, Hefei 230026, People's Republic of China\\
% ...
$^{81}$ University of Turin and INFN, (A)University of Turin, I-10125, Turin, Italy; (B)University of Eastern Piedmont, I-15121, Alessandria, Italy; (C)INFN, I-10125, Turin, Italy\\
$^{82}$ Uppsala University, Box 516, SE-75120 Uppsala, Sweden\\
% ...
$^{87}$ Zhengzhou University, Zhengzhou 450001, People's Republic of China\\

\vspace{0.2cm}
$^{\dagger}$ Deceased\\
% ...
$^{c}$ Also at the Novosibirsk State University, Novosibirsk, 630090, Russia\\
% ...
$^{h}$ Also at State Key Laboratory of Nuclear Physics and Technology, Peking University, Beijing 100871, People's Republic of China\\
% ...

}
%% ends here %%

\end{center}
\end{small}
\vspace{0.4cm}
}

\vspace{4cm}

\date{\it \small \bf \today}

\begin{abstract}
Using a sample of $e^{+}e^{-}$ annihilation data corresponding to an integrated luminosity of 4.4 $\rm{fb}^{-1}$ collected with the BESIII detector at the BEPCII collider and produced at center-of-mass energies from $4600$ to $4698~\rm{MeV}$, 
an amplitude analysis is performed of the singly Cabibbo-suppressed decay $\Lambda^{+}_{c}\to pK^{+}K^{-}$.
The branching fractions of $\Lambda^{+}_{c}\to p\phi(1020)$, $pf_{0}(980)$, $\Lambda(1405)K^{+}$, and $\Lambda(1670)K^{+}$ are measured, where the latter two modes are decays that are observed for the first time.
At the same time, with the detection efficiency based on the results of the amplitude analysis, the branching fraction of $\Lambda^{+}_{c}\to pK^{+}K^{-}$ is updated to be $(9.94\pm0.65_{\text{stat.}}\pm0.50_{\text{syst.}})\times10^{-4}$, which is consistent with the current world average value within one standard deviation. The result supersedes the previous BESIII measurement with precision improved by approximately a factor of 1.5.
\end{abstract}

%\pacs{13.30.Ce, 12.15.Ji}

\maketitle

\oddsidemargin  -0.2cm
\evensidemargin -0.2cm

\section{Introduction}\label{sec:intro}

Charmed baryons were first observed in the 1970s, with the lightest one, $\Lambda^+_c$, sparking extensive research into the properties and structure of the charmed baryon family~\cite{Abrams:1979iu}.
Distinct from the subsequently discovered $\Xi_c^{+,0}$ and $\Omega_c^0$ baryons, the $\Lambda^+_c$ provides a comparatively simple system for exploring the dynamics of charmed baryons~\cite{Li:2025nzx}. It also offers crucial insights into the structure and decay mechanisms of doubly charmed baryons such as $\Xi_{cc}$ and $\Omega_{cc}$~\cite{Yu:2017zst}.

Quantum chromodynamics (QCD)~\cite{Politzer:1973fx,Gross:1973id}, the fundamental theory describing the strong interaction among quarks and gluons, plays a central role in predicting the decay of quarks bound within charmed baryons. Among the various decay categories, singly Cabibbo-suppressed (SCS) decays are of particular interest because of their sensitivity to charge–parity ($\textit{C\!P}$) violation. These decays can carry a significant weak phase, thereby providing enhanced sensitivity to possible $\textit{C\!P}$-violating effects beyond the Standard Model~\cite{Wang:2024wrm,BESIII:2025zbz}.
The SCS decay $\Lambda^{+}_{c} \to pK^{+}K^{-}$ is regarded as a golden channel for investigating $\textit{C\!P}$ violation in charmed baryon decays, owing to its relatively large branching fraction and favorable detection efficiency. A comprehensive study of the intermediate resonant contributions in this decay is vital for improving our understanding of the underlying decay dynamics and for refining searches for $\textit{C\!P}$ violation. Moreover, the use of high-precision signal models allows for a more accurate description of signal behavior in experimental data, which is essential for reliably identifying and quantifying possible $\textit{C\!P}$-violating effects in baryonic systems.

\begin{figure}[htbp]
	\centering
\subfigure[]{\includegraphics[width=0.48\linewidth]{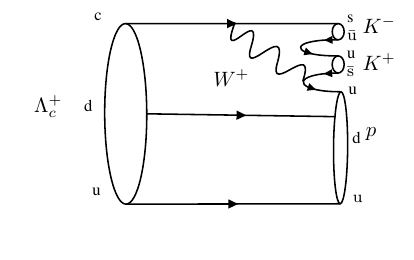}\label{fig:feynman:sub_a}}
\subfigure[]{\includegraphics[width=0.48\linewidth]{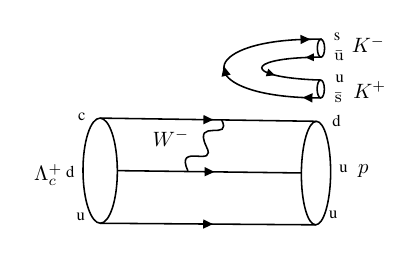}\label{fig:feynman:sub_b}}
\subfigure[]{\includegraphics[width=0.48\linewidth]{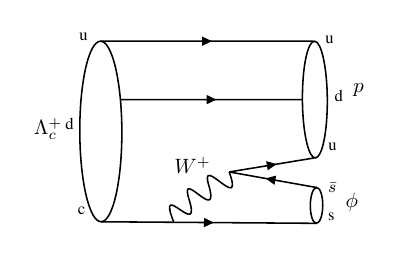}\label{fig:feynman:sub_c}}
	\caption{
Feynman diagrams of $\Lambda^{+}_{c}\to pK^{+}K^{-}$ via internal W-emission (a), 
$\Lambda^{+}_{c}\to pK^{+}K^{-}$ via W-exchange (b), 
and $\Lambda^{+}_{c}\to p\phi(1020)$ via internal W-emission (c).
}
	\label{fig:feynman}
\end{figure}

Situated at the interface between perturbative and non-perturbative QCD regimes, the charm quark’s role in this decay has attracted considerable theoretical attention, particularly in the $\Lambda^+_c \to p\phi(1020)$ channel, which provides the leading narrow-vector contribution to $\Lambda^+_c \to pK^{+}K^{-}$. 
The corresponding Feynman diagrams for $\Lambda^+_c \to pK^{+}K^{-}$ and $\Lambda^+_c \to p\phi(1020)$ are shown in Fig.~\ref{fig:feynman}. The $\Lambda^+_c \to p\phi(1020)$ process is comparatively simple, proceeding primarily through the factorizable diagram (Fig.~\ref{fig:feynman:sub_c})~\cite{Cheng:1993gf}. However, theoretical predictions for this channel differ significantly among models~\cite{Korner:1992wi,Zenczykowski:1993hw,Zenczykowski:1993jm,Datta:1995mn,Cheng:1995fe,Ivanov:1997ra}, as summarized—--together with available experimental measurements--—in Table~\ref{int:restable}. 

Beyond absolute branching-fraction estimates for the $p\phi(1020)$ mode, the substructure of $\Lambda_c^+ \to pK^+K^-$ has also attracted theoretical interest. In particular, a chiral-unitary study suggested that~\cite{Wang:2020pem}, in addition to the narrow $\phi(1020)$ peak, the $K^+K^-$ spectrum may contain a near-threshold scalar enhancement associated mainly with $f_0(980)$. 
Similar to the LHCb amplitude analysis of the $\Lambda_c^+ \to pK^-\pi^+$ decay~\cite{LHCb:2022sck}, we employ analogous amplitude analysis techniques to determine more precisely the relative contributions of different components in the $\Lambda_c^+\to pK^+K^-$ system. This approach also enables a quantitative treatment of possible nonresonant amplitudes and interference effects, thereby providing useful insights into the underlying dynamical mechanisms of the decay.
The amplitude analysis provides a powerful tool for disentangling the contributing intermediate states and their interferences, offering more profound insight into the decay dynamics of charmed baryons. In particular, it allows the identification of structures in the $pK^{-}$ invariant mass spectrum, thereby refining the modeling of signal components and reducing systematic uncertainties in the branching fraction measurement.

\begin{table*}[htbp]
	\centering
	\caption{Predictions for and measurements of branching fractions for $\Lambda_c^+\to pK^{+}K^{-}$ and $\Lambda_c^+\to p\phi(1020)$. The superscript “*” indicates a range of predicted values that depend on the renormalization scheme used in the corresponding theoretical model.}
	\label{int:restable}
	\begin{tabular}{c @{\hspace{1cm}} c @{\hspace{1cm}} c}
		\hline\hline
		Prediction or measurement                             & $\Lambda_c^+\to pK^{+}K^{-}$~(\%) & $\Lambda_c^+\to p\phi(1020)$~(\%)   \\
		\hline\hline
            K{\"o}rner(1992), CCQM~\cite{Korner:1992wi}             & ...   & $0.003$  \\
            \hline
		{\'Z}encaykowski(1994), Pole~\cite{Zenczykowski:1993hw} & ...   & $0.10$   \\
		\hline
            {\'Z}encaykowski(1994), Pole~\cite{Zenczykowski:1993jm} & ...   & $0.14$   \\
            \hline
		Alakabha Datta(1995), CA~\cite{Datta:1995mn}            & ...   & $0.063$ \\
		\hline
            Cheng, Tseng(1993), Pole~\cite{Cheng:1995fe}            & ...   & $0.073$  \\
            \hline
            Ivanov(1998), CCQM~\cite{Ivanov:1997ra}                 & ...   & $0.138-0.657$* \\
		\hline\hline
            CLEO(1996)~\cite{CLEO:1995moc}                          & $0.244\pm0.056\pm0.044$     & $0.150\pm0.038\pm0.019$ \\
            \hline
            Belle(2002)~\cite{Belle:2001hyr}                        & $0.088\pm0.013\pm0.013$     & $0.094\pm0.013\pm0.013$ \\
            \hline
            BESIII(2016)~\cite{BESIII:2016ozn}                      & $0.107\pm0.016\pm0.007$     & $0.106\pm0.019\pm0.010$ \\
            \hline
            LHCb(2018)~\cite{LHCb:2017xtf}                          & $0.106\pm0.002\pm0.005$     & ... \\
            \hline
		PDG(2024)~\cite{ParticleDataGroup:2024cfk}                        & $0.106\pm0.006$             & $0.106\pm0.014$ \\
		\hline\hline
	\end{tabular}
\end{table*}

Using 4.4~$\rm{fb}^{-1}$ of $e^{+}e^{-}$ annihilation data collected at the center-of-mass (CM) energies 4600, 4628, 4641, 4661, 4682, and 4698~$\rm{MeV}$ with the BESIII detector at the BEPCII collider~\cite{Ke:2023qzc}, we perform the first amplitude analysis of the decay $\Lambda_{c}^{+}\to pK^{+}K^{-}$ and determine its branching fraction with improved precision.

\section{BESIII detector and Monte Carlo simulation}

The BESIII detector~\cite{Ablikim:2009aa} records symmetric $e^+e^-$ collisions 
provided by the BEPCII storage ring~\cite{Yu:IPAC2016-TUYA01}
in the CM energy ranging from 1.84 to 4.95~GeV,
with a peak luminosity of $1.1 \times 10^{33}\;\text{cm}^{-2}\text{s}^{-1}$ 
achieved at $\sqrt{s} = 3.773\;\text{GeV}$. BESIII has collected large data samples in this energy region~\cite{Ablikim:2019hff}. The cylindrical core of the BESIII detector covers 93\% of the full solid angle and consists of a helium-based multilayer drift chamber~(MDC), a plastic scintillator time-of-flight system~(TOF), and a CsI(Tl) electromagnetic calorimeter~(EMC), which are all enclosed in a superconducting solenoidal magnet providing a 1.0~T magnetic field. The solenoid is supported by an octagonal flux-return yoke with resistive plate counter muon identification modules interleaved with steel. The charged-particle momentum resolution at $1~{\rm GeV}/c$ is $0.5\%$, and the specific ionization energy loss in the MDC (${\rm d}E/{\rm d}x$) resolution is $6\%$ for electrons from Bhabha scattering. The EMC measures photon energies with a
resolution of $2.5\%$ ($5\%$) at $1$~GeV in the barrel (end cap)
region. The time resolution in the plastic scintillator TOF barrel region is 68~ps, while
that in the end cap region is 110~ps. The end cap TOF
system was upgraded in 2015 using multigap resistive plate chamber
technology, providing a time resolution of 60~ps~\cite{Cao:2020ibk}.

Monte Carlo (MC) simulated data samples produced with a {\sc
geant4}-based~\cite{geant4} software package, which
includes the geometric description of the BESIII detector and the
detector response, which are used to determine detection efficiencies
and to estimate backgrounds. The simulation models the beam
energy spread and initial state radiation (ISR) in the $e^+e^-$
annihilations with the generator {\sc
kkmc}~\cite{ref:kkmc}.
The inclusive MC sample includes the production of open charm
processes, the initial state radiation production of vector charmonium(-like) states,
and the continuum processes incorporated in {\sc
kkmc}~\cite{ref:kkmc}.
All particle decays are modeled with {\sc
evtgen}~\cite{ref:evtgen} using branching fractions 
either taken from the
Particle Data Group (PDG)~\cite{ParticleDataGroup:2024cfk}, where available,
or otherwise estimated with {\sc lundcharm}~\cite{ref:lundcharm}.
Final state radiation
from charged final state particles is incorporated using the {\sc
photos} package~\cite{photos2}.
The signal MC samples for amplitude analysis and branching fraction measurement are generated using different models. For the amplitude analysis, a phase space (PHSP) MC sample is used in the MC integration. While for the branching fraction measurement, the signal MC sample is generated based on the result of amplitude analysis for a more accurate efficiency.

\section{Event selection}
%%%%%%    Charged track     %%%%%%%%%%
Charged tracks detected in the MDC are required to satisfy $|\!\cos\theta|<0.93$, 
where $\theta$ is the polar angle with respect to the $z$-axis, defined as the symmetry axis of the MDC. 
Each track must have a distance of closest approach to the interaction point (IP) of less than 10~cm along the beam direction ($V_{z}$) and less than 1~cm in the transverse plane ($V_{r}$). Particle identification (PID) for charged tracks combines measurements of the ionization energy loss (${\rm d}E/{\rm d}x$) in the MDC and the flight time measured by the TOF system. 
For each charged particle hypothesis $h = (p, K, \pi)$, a corresponding likelihood $\mathcal{L}(h)$ is calculated. 
A track is identified as a proton if $\mathcal{L}(p)>\mathcal{L}(K)$ and $\mathcal{L}(p)>\mathcal{L}(\pi)$. 
Similarly, charged kaons and pions are identified according to the larger likelihood between the kaon and pion hypotheses, i.e., $\mathcal{L}(K)>\mathcal{L}(\pi)$ or $\mathcal{L}(\pi)>\mathcal{L}(K)$, respectively.

%%%%%%     Further vertex cut for proton     %%%%%%%%%%
An additional requirement of $V_{r}<0.2$~cm is imposed on proton candidates to suppress background protons originating from beam-related interactions or secondary processes. Such backgrounds may arise from interactions of the electron/positron beams or final-state particles with the inner wall of the MDC or residual gas inside the beam pipe, as discussed in Ref.~\cite{BESIII:2016ozn}.

%%%%%%    $\Lambda_{c}$ Reconstruction   %%%%%%%%%%

With the preliminarily selected proton and kaon candidates, $\Lambda_c^{+}$ signal candidates are reconstructed from all possible combinations of $pK^{+}K^{-}$. 
Two kinematic variables related to energy and momentum conservation are used to identify $\Lambda_c^{+}$ candidates.
The first variable is the energy difference, defined as 
$\Delta E = E - E_{\rm{beam}}$, 
where $E$ is the total reconstructed energy of the $\Lambda_c^{+}$ candidate in the $e^{+}e^{-}$ CM system and $E_{\rm{beam}}$ is the calibrated beam energy. 
The $\Lambda_c^{+}$ signal is expected to peak near zero in the $\Delta E$ distribution. 
The second variable is the beam-energy-constrained mass, 
$M_{\rm{BC}} \equiv \sqrt{E_{\rm{beam}}^{2}/c^{4} - |\vec{p}\,|^{2}/c^{2}}$, 
where $\vec{p}$ is the reconstructed total momentum of the $\Lambda_c^{+}$ candidate in the CM frame. 
To obtain a clean sample of $\Lambda_c^{+}$ candidates, loose preselection requirements are applied: 
$|\Delta E| < 0.1~\rm{GeV}$ and $M_{\rm{BC}} > 2.25~\rm{GeV}/c^{2}$. 
If multiple candidates satisfy these conditions within a single event, the candidate with the smallest $|\Delta E|$ is retained. 
Owing to the fact that all final-state particles are charged tracks and to the low-background $e^+e^-$ collision environment at BESIII, the multiple-candidate fractions are below $1\%$ at all energy points. Moreover, good agreement is observed between data and MC; therefore, no systematic uncertainty associated with multiple candidates is assigned.
The selection of the optimal $\Delta E$ window is further refined by maximizing the Figure of Merit (FOM), 
defined as 
${\rm FOM} = \frac{S}{\sqrt{S+B}} \times \frac{S}{S+B}$~\cite{LHCb:2021chn,LHCb:2022ogu}, 
where $S$ and $B$ denote the numbers of signal and background events, respectively, estimated from inclusive background MC samples. 
For the amplitude analysis, a stringent requirement of $|\Delta E| < 0.005~\rm{GeV}$ is imposed, 
while for the branching fraction measurement, a looser selection of $|\Delta E| < 0.02~\rm{GeV}$ is applied. 
To evaluate the residual combinatorial background that survives the selection criteria, 
an unbinned maximum-likelihood fit to the $M_{\rm{BC}}$ distribution is performed separately for each energy point.
In the fit, the signal is modeled using an MC-simulated shape convolved with a Gaussian function to account for small discrepancies between data and simulation due to imperfect detector modeling and beam-energy spread. The parameters of the Gaussian function are floated in the fit, with the fitted mean and width both reaching the level of $0.1~\mathrm{MeV}/c^{2}$, indicating that the MC simulation provides a good description of the data.
The background is described by an ARGUS function~\cite{ARGUS:1990hfq}, defined as
\begin{equation}
    f(M_{\rm{BC}},E_{0},c,p) = 
    M_{\rm{BC}}
    \Bigg(1 - \Big(\frac{M_{\rm{BC}}}{E_{0}}\Big)^{2}\Bigg)^{p}
    \times
    e^{\,c\cdot\big(1 - \frac{M_{\rm{BC}}}{E_{0}}\big)^{2}},\notag
\end{equation}
where $E_{0}$ is the endpoint of $M_{\rm{BC}}$ fixed to the beam energy, 
$p$ is the power parameter (fixed to $0.5$ by default), 
and $c$ is a free scale factor. 
The total probability density function (PDF) is expressed as the sum of the signal and the background components.

Based on the fits to the $M_{\rm BC}$ distributions of data at each individual energy point, the background-shape parameters obtained from the corresponding fit are fixed in the subsequent analyses that employ the $\hbox{$_s$}{\cal P}lot$ technique~\cite{Pivk:2004ty}.
This method assigns event-by-event statistical weights, known as sWeights, 
which are derived as a function of a discriminating variable ($M_{\rm{BC}}$). 
These weights are constructed such that the signal distribution is statistically normalized, while being orthogonal to the background distribution. 
Consequently, after applying the sWeights, background contributions are effectively removed from the signal distribution. 
The resulting weighted sample is used for the amplitude analysis to ensure a high-purity signal dataset. 
By combining data from all energy points, a total of $353 \pm 22$ signal events is obtained.

\section{Amplitude Analysis}
Once all selection criteria have been applied, the remaining signal events are used for the amplitude analysis. 
This analysis is performed using the helicity amplitude formalism~\cite{Chung:1971ri,Richman:1984gh}, 
implemented within the open-source framework \textsc{TF-PWA}~\cite{tfpwa}.

For a generic two-body decay process of the form $1 \to 2 + 3$, 
the helicity amplitude can be expressed as
\begin{equation}
A^{1\to 2 + 3}_{\lambda_{1},\lambda_{2},\lambda_{3}} = 
H^{1\to 2 + 3}_{\lambda_{2},\lambda_{3}} 
D^{J_{1}}_{\lambda_{1},\lambda_{2}-\lambda_{3}}(\phi,\theta,0),\notag
\end{equation}
where $J$ and $\lambda$ denote the spin and helicity of the respective particles. 
The amplitude $H^{1\to 2 + 3}_{\lambda_{2},\lambda_{3}}$ is parameterized using the LS coupling formalism~\cite{Chung:1997jn}, 
which includes the Blatt–Weisskopf barrier factor~\cite{VonHippel:1972fg}. 
In this framework, the LS coupling strengths serve as the direct fit parameters of the analysis.

\begin{figure}[htbp]\centering
	\includegraphics[width=0.9\linewidth]{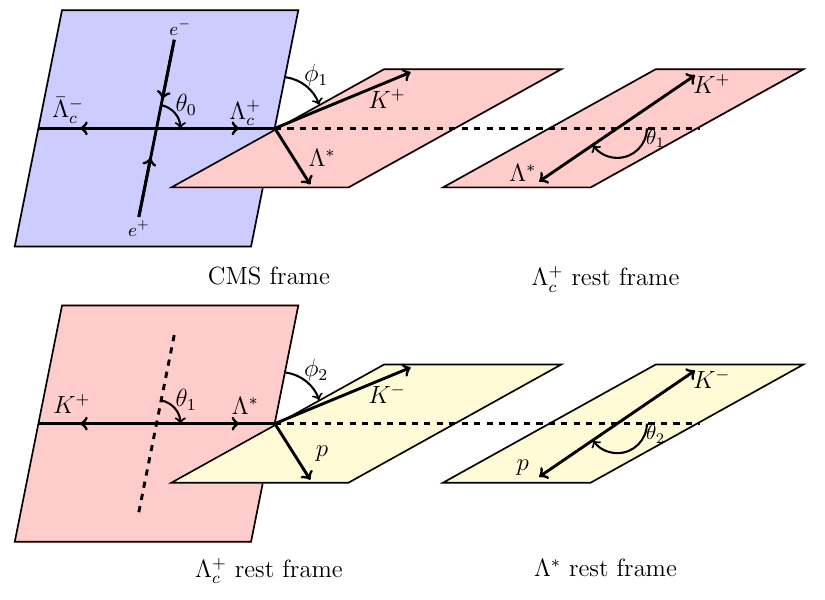}
	\caption{Definition of the helicity frame for $e^{+}e^{-}\to\Lambda^{+}_{c}\bar{\Lambda}^{-}_{c}, \Lambda^{+}_{c}\to \Lambda^{*}K^{+}$.}
	\label{fig:topo}
\end{figure}

As an illustrative example, consider the decay chain 
$e^{+}e^{-} \to \Lambda^{+}_{c}\,\bar{\Lambda}^{-}_{c}$, 
followed by $\Lambda^{+}_{c} \to \Lambda^{*}K^{+}$. 
The corresponding helicity angles and cascade decay topology are depicted in Fig.~\ref{fig:topo}. 
In the Wigner $D$-function $D^{J_{1}*}_{\lambda_{1},\lambda_{2}-\lambda_{3}}(\phi,\theta,0)$, 
the variables $\phi$ and $\theta$ denote the helicity angles defined as shown in Fig.~\ref{fig:topo}.

The amplitudes are constructed from $e^{+}e^{-}$ to the final two-body decays. The parameters describing the amplitude from $\Lambda^{+}_{c}$ decays to the final two-body state are shared for all energy points. 

The full amplitude for each component is the product of every two body decay amplitude and the corresponding dynamic part. For example, in the $\Lambda^{+}_{c}\to pK^+K^-$ decay, the amplitude of the $pK^-(\Lambda^*)$ resonance is defined as
\begin{equation}
A_{\lambda_{\gamma^{*}},\lambda_{p}}^{\Lambda^*}=\sum_{\lambda_{\bar{\Lambda}^{-}_{c}}}A^{\gamma^{*}\to\Lambda^{+}_{c}\bar{\Lambda}^{-}_{c}}_{\lambda_{\gamma^{*}},\lambda_{\Lambda^{+}_{c}},\lambda_{\bar{\Lambda}^{-}_{c}}}A^{\Lambda^{+}_{c}\to\Lambda^*K^+}_{\lambda_{\Lambda^{+}_{c}},\lambda_{\Lambda^*},0}R_{\Lambda^*}(M_{pK^-})A^{\Lambda^*\to pK^-}_{\lambda_{\Lambda^*},\lambda_{p},0}\notag.
\label{PWA:amp}
\end{equation}

To enhance the precision of the $\Lambda^{+}_{c}$ decay amplitude measurement, the decay parameters of 
$e^{+}e^{-}\to\Lambda^{+}_{c}\bar{\Lambda}^{-}_{c}$ are fixed according to the BESIII measurement~\cite{BESIII:2023rwv}. For the $\Lambda_{c}^{+}$ decay, the three-body decay is described by several quasi-two-body sequential two-step decays. The amplitude for a full decay chain contains a propagator $R$~\cite{BESIII:2022udq}, using different models for different resonances. For the $\Lambda(1670)$ and $\phi(1020)$ resonances, their line shapes are described with the relativistic Breit-Wigner formula~\cite{BESIII:2022udq} 
\begin{equation*}
    R(m)=\frac{1}{m^{2}_{0}-m^{2}-im_{0}\Gamma(m)},
\end{equation*}
where $\Gamma(m)$ is the mass-dependent width. The nominal mass and width of $\Lambda(1670)$ are taken from the PDG~\cite{ParticleDataGroup:2024cfk} while they are free parameters for $\phi(1020)$. For $\Lambda(1405)$ and $f_{0}(980)$, their line shapes are described with a two-coupled-channels Flatt\'{e} model~\cite{Flatte:1976xu} 
\begin{equation*}
  R(m)=\frac{1}{m^{2}_{0}-m^{2}-i(g_{1}\rho_{1}+g_{2}\rho_{2})}
\end{equation*}
with phase-space factors $\rho_k$ and coupling strengths $g_k$. 
For $f_{0}(980)$, channels 1 and 2 denote the coupled channels $\pi\pi$ and $KK$~\cite{BES:2004twe}, while they denote $pK$ and $\Sigma\pi$ for $\Lambda(1405)$~\cite{LHCb:2015yax}. 

Finally, the total amplitude is derived by summing the amplitudes of all possible resonances as
\begin{equation*}
\begin{aligned}
\mathcal{A}^{total}_{\lambda_{\gamma^{*}},\lambda_{p}}
&=A^{\phi}_{\lambda_{\gamma^{*}},\lambda_{p}}
+ A^{f_{0}(980)}_{\lambda_{\gamma^{*}},\lambda_{p}}
\\[6pt]
&\quad
+ \sum_{\lambda'_{p}}
\Bigl(\sum A^{\Lambda^*}_{\lambda_{\gamma^{*}},\lambda'_{p}}\Bigr)
D^{1/2}_{\lambda'_{p},\lambda_{p}}(\alpha_{p},\beta_{p},\gamma_{p})
\end{aligned}
\end{equation*}
where the extra $D$-function is added to align the spins of the final-state protons from different decay chains.

The construction of the signal PDF and the derivation of the fit fraction (FF), interference, and the corresponding statistical uncertainties follow the methodologies employed in the previous BESIII measurement~\cite{BESIII:2022udq}. Taking $\Lambda(1670)$ as an example, the FF is defined to be 
\begin{equation*}
    \mathrm{FF}=\frac{\mathcal{B}(\Lambda^{+}_{c}\to\Lambda(1670)K^{+}, \Lambda(1670)\to pK^{-})}{\mathcal{B}(\Lambda^{+}_{c}\to pK^{+}K^{-})}.
\end{equation*}

The negative log likelihood (NLL) is calculated as the sum over all signal candidates, weighted by the sWeight factor $w_i$~\cite{Pivk:2004ty}, 
\begin{equation*}
    -\ln L=-a \sum_{i\in\mathrm{data}} w_i \ln P(p_i),
\end{equation*}
with the normalization factor $a = \frac{\sum_{i\in\mathrm{data}} w_i}{\sum_{i\in\mathrm{data}} w_i^2}$~\cite{Langenbruch:2019nwe}.

To establish the baseline solution for the amplitude analysis, a statistical significance test is performed for each possible intermediate component. 
Distinct structures corresponding to $\phi(1020)$, $\Lambda(1405)$, and $\Lambda(1670)$ are clearly observed in the $K^{+}K^{-}$ and $pK^{-}$ invariant mass spectra; hence, these components are incorporated into the baseline model. 
The statistical significance of each component is evaluated by comparing the change in the negative log-likelihood (NLL) between fits with and without the component, while accounting for the associated change in the number of degrees of freedom. 
As a result, the baseline solution includes the $\phi(1020)$, $f_{0}(980)$, $\Lambda(1405)$, and $\Lambda(1670)$ resonant components, with no evidence for a significant nonresonant contribution. 
For the $\phi(1020)$ resonance, its intrinsic width is comparable to the detector resolution; therefore, resolution effects must be taken into account. 
These effects are incorporated into the probability density function (PDF) by convolving the line shape with a detector resolution function. 
The resolution parameters are obtained by fitting the $K^{+}K^{-}$ invariant mass spectrum with a Breit–Wigner line shape convolved with a double-Gaussian resolution model.
To accurately describe the resolution effects in the $\Lambda_c^{+} \to pK^{+}K^{-}$ decay, the relationship between the true and reconstructed momenta is integrated into the model. 
The resulting PDF is expressed as a weighted sum of amplitude squares, modulated by detector effects, providing a precise description of the decay dynamics and an effective treatment of the experimental resolution.

\begin{figure*}[htbp]
	\centering
	\includegraphics[width=0.32\textwidth]{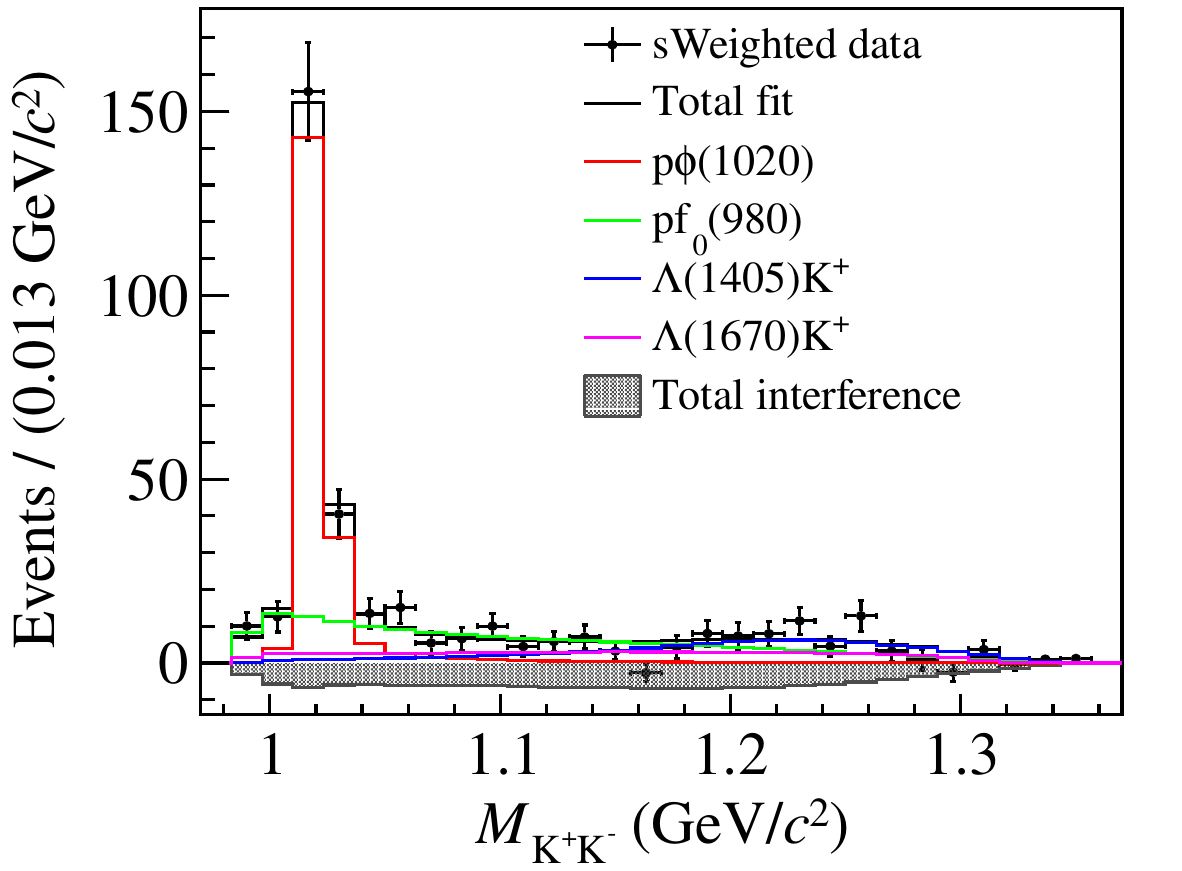}
	\includegraphics[width=0.32\textwidth]{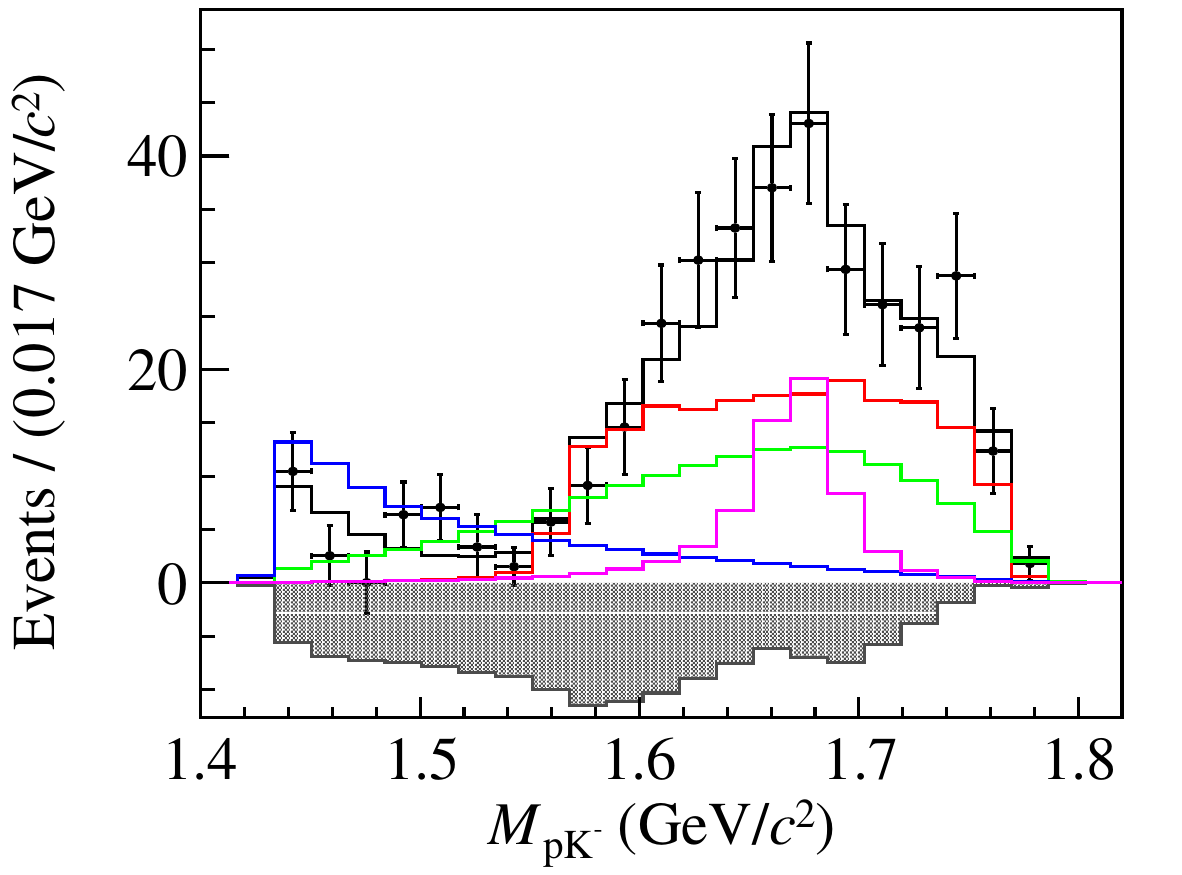}
	\includegraphics[width=0.32\textwidth]{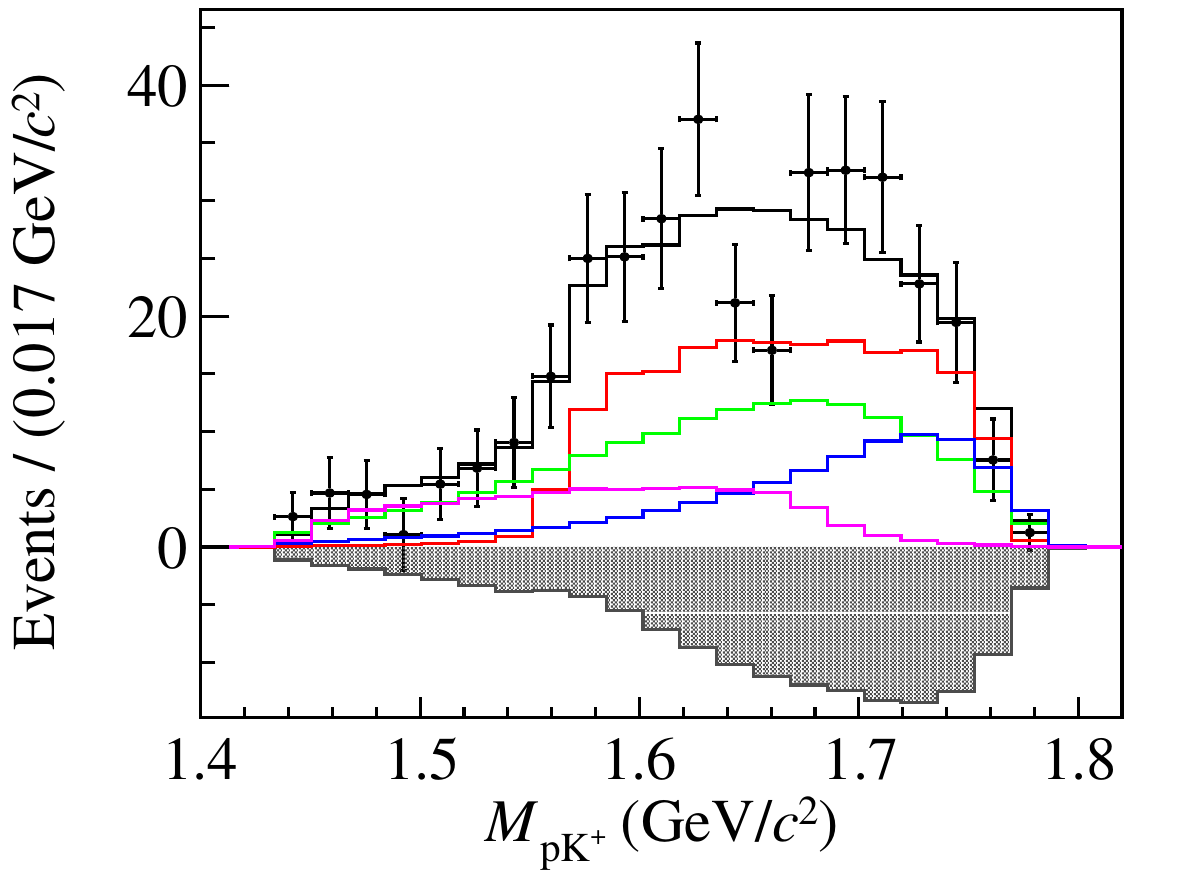}
	\caption{Projections of the amplitude fit results on the $M_{K^{+}K^{-}}$, $M_{pK^{-}}$, and $M_{pK^{+}}$ spectra. The points with error bars are weighted data at all energy points. Different colored curves represent different components, as shown in the legend.\label{fig:fit_mass}}
\end{figure*}

The projections of the fit results on the invariant mass spectra are shown in Fig.~\ref{fig:fit_mass}, and the obtained fit fractions (FFs) together with their statistical significances are summarized in Table~\ref{tab:finalresults}. 
To validate the stability and reliability of the fit procedure, an input–output (IO) test is performed using 500 pseudoexperiments (“toy” samples), each containing the same number of events as the data. 
Each toy sample includes both signal and background contributions: the signal is generated using the central values from the nominal fit, while the background is modeled using inclusive and hadronic MC samples. 
The effects of the detector efficiency are appropriately incorporated into each simulated sample. 
The fitted FFs and their statistical uncertainties from the nominal fit are then corrected based on the pull distributions obtained from the IO test.

%%%%%%%%%%%%%%%%%%
\begin{table}[htbp]
\begin{center}
\caption{The FFs and statistical significances~($\mathcal S$) for different components in the baseline solution. The total FF is 130\%. The first uncertainty is statistical, and the second is systematic.} 
\small
	$\begin{array}{ccc}
	\hline \hline 
	\text { Process } & \text{FF (\%)} & \mathcal S  \\
	\hline 
    p\phi(1020)        & 57\pm5\pm2 &16.6\sigma  \\
    pf_{0}(980)        & 40\pm16\pm15 & 3.6\sigma  \\
    \Lambda(1405)K^{+} & 21\pm9\pm6 & 4.6\sigma  \\
	\Lambda(1670)K^{+} & 12\pm10\pm8 & 5.0\sigma \\
	\hline \hline
	\end{array}$
\label{tab:finalresults}
\end{center}
\end{table}
%%%%%%%%%%%%%%%%%%

\section{Branching Fraction Measurement}
\begin{figure}[htbp]\centering
	\includegraphics[width=0.9\linewidth]{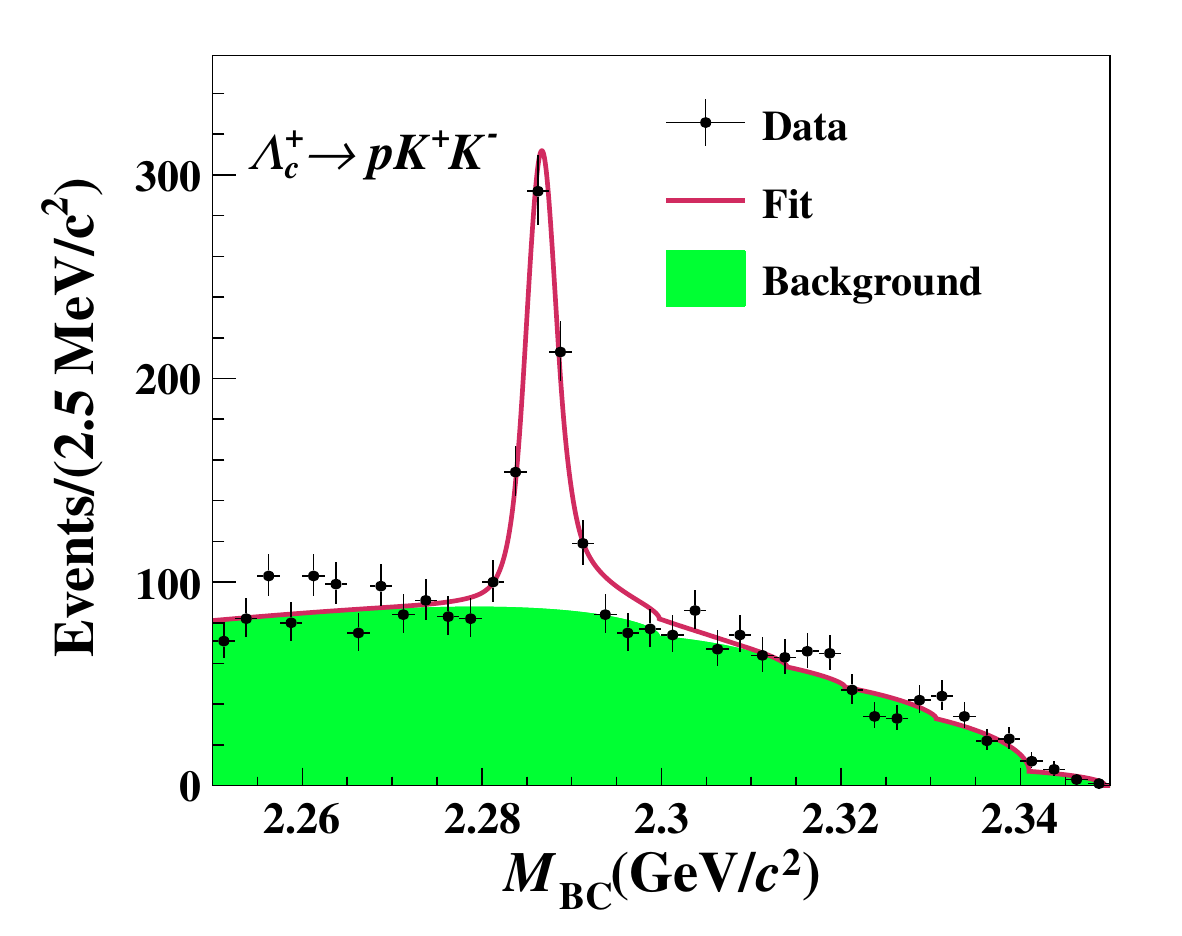}
	\caption{
    Fit results to the $M_{\rm BC}$ distributions of the accepted candidates for
    $\Lambda_c^+\to pK^+K^-$, obtained by summing the individual fit plots from
    all energy points. The black dots with error bars represent the summed data,
    and the green-shaded region indicates the summed combinatorial background.
    The requirement $|\Delta E| < 0.02~{\rm GeV}$ is applied in this figure.}
	\label{fig:mBC_4600}
\end{figure}

Benefiting from the amplitude analysis technique, the MC simulation can be further improved, thereby enabling a more precise measurement of the branching fraction of $\Lambda_c^+ \to pK^+K^-$. As mentioned earlier, a relaxed requirement of $|\Delta E| < 0.02~\rm{GeV}$ is applied in this part of the study.
In the nominal fit, a simultaneous fit to the $M_{\rm BC}$ distributions at six center-of-mass energies is carried out, with a common branching fraction shared among all energy points. 
The fit to the $M_{\rm{BC}}$ distribution of the accepted candidates for $\Lambda_c^+ \to pK^+K^-$ combined over all energy points is shown in Fig.~\ref{fig:mBC_4600}. The signal component is described by the signal MC shape derived from the amplitude analysis, convolved with a Gaussian resolution function whose parameters are left free in the fit. 
At each energy point, 100,000 signal MC events are generated to model the signal shape. 
The Gaussian parameters are constrained to be the same across all energy points to ensure a consistent description of the resolution. 
The background is modeled with an ARGUS function~\cite{ARGUS:1990hfq}, where the endpoint parameter is fixed to the beam energy $E_{\rm beam}$ corresponding to each energy point.
The detection efficiency is determined from signal MC samples. 
For each energy point, a dedicated sample of 1,000,000 simulated events is generated, in which one of the $\Lambda^{+}_{c}(\bar{\Lambda}^{-}_{c})$ baryons decays into the signal mode and the other into all possible final states. 
The detection efficiency is obtained by counting the number of reconstructed signal events that satisfy all selection criteria within the signal region. 
The efficiencies determined for the six energy points are summarized in Table~\ref{data:eff}.

\begin{table}[htbp]
\begin{center}
\small
     \caption{Detection efficiencies for each decay mode at six energy points. The uncertainties on the efficiencies are due to the finite size of the simulated samples.}
	\label{data:eff}
	\begin{tabular}{c  c  c  c  c  }

		\hline\hline
		$\sqrt{s} (\rm{MeV})$ &  & $\varepsilon_i(\%)$    &  & $N^i_{\rm sig}$       \\
		\hline
		$4600 $ &  & $39.55\pm0.05$               &  &  $76.9\pm5.0$           \\
		\hline
		$4628 $ &  & $36.62\pm0.05$               &  &  $65.2\pm4.3$           \\
		\hline
		$4641 $ &  & $36.11\pm0.05$               &  &  $68.9\pm4.5$           \\
		\hline
		$4661 $ &  & $35.10\pm0.05$               &  &  $65.9\pm4.3$           \\
		\hline
		$4682 $ &  & $34.08\pm0.05$               &  &  $193.0\pm12.6$         \\
		\hline
		$4698 $ &  & $33.33\pm0.05$               &  &  $55.7\pm3.6$           \\
		\hline\hline
	\end{tabular}
\end{center}
\end{table}

The branching fraction is determined by
\begin{equation*}
    \mathcal{B}=\frac{N^i_{\rm sig}}{2\cdot[N_{(\Lambda^{+}_{c}\bar{\Lambda}^{-}_{c})_{i}}\cdot\varepsilon_{i}]}, (i=1,6),
\end{equation*}
where $N_{(\Lambda^{+}_{c}\bar{\Lambda}^{-}_{c})_i}$ is number of the $\Lambda^{+}_{c}\bar{\Lambda}^{-}_{c}$ pairs at the $i$-th energy point, taken from Ref.~\cite{BESIII:2026qbp}.
The branching fraction for $\Lambda^{+}_{c}\to pK^{+}K^{-}$ is a fit parameter. 
Finally, the fit returns a signal yield of $526\pm34$ events, and the branching fraction of $\Lambda^{+}_{c}\to pK^{+}K^{-}$ is measured to be $(9.94\pm0.65)\times10^{-4}$.

\section{Systematic uncertainties}
In the amplitude analysis, systematic uncertainties arise from several sources, including resonance parameters, the radius parameter, background size, background shape, resonant components, polarization parameters, and the discrepancy between data and MC simulations.

\begin{table*}[!htbp]
    \begin{center}
  \caption{The systematic uncertainties of the amplitude analysis.}
  \small
  \begin{tabular}{lcccc}
      \hline \hline
			Source            &  $\mathrm{FF}_{\phi(1020)}~(\%)$ & $\mathrm{FF}_{f_{0}(980)}~(\%)$ &$\mathrm{FF}_{\Lambda(1405)}~(\%)$&  $\mathrm{FF}_{\Lambda(1670)}~(\%)$ \\
			\hline
			Resonance parameter 		       	&0.8	&34.8	&23.0&39.8	\\
			Radius parameter   					&2.1	&1.8	&4.3  &2.4	\\
			Background size	            		&0.5	&10.2	&4.7  &6.8	\\
			Background shape					&0.7	&1.7	&7.5  &13.3	\\
                Resonant component                  &1.0   &16.4  &19.2 &13.3   \\
                Polarization parameter              &0.3   &3.2   &2.6  &7.0    \\
			Data/MC difference 					&0.6	&1.2	&1.9  &0.6	\\
			\hline
		    Total             					&2.7	&40.0	&31.7 &45.1	\\
      \hline \hline
    \end{tabular}
    \label{tab:sys_err:pwa}
    \end{center}
\end{table*}

The systematic uncertainties associated with the resonance parameters originate from the fixed values used to describe each resonance. 
To estimate these uncertainties, Gaussian sampling is performed using the central values and standard deviations of the fixed parameters. 
For each sampled set, the amplitude fit is repeated, and the resulting distribution of each FF is fitted with a Gaussian function; the corresponding standard deviation is taken as the systematic uncertainty.
Systematic uncertainties related to the Blatt–Weisskopf barrier radius parameter are evaluated by varying $p_{r}=1/d$ within the range $0.17~\rm{GeV}/c$ to $0.37~\rm{GeV}/c$, as recommended in Ref.~\cite{BESIII:2019dme}. 
The uncertainty due to the background yield is estimated by varying the background normalization by $\pm1\sigma$ in the $\hbox{$_s$}{\cal P}lot$ procedure and repeating the sWeight extraction for the amplitude analysis. 
For the background shape, the systematic effect is assessed by replacing the nominal ARGUS function with the background shape derived from MC simulation in the $\hbox{$_s$}{\cal P}lot$ procedure, followed by a re-extraction of the sWeights and repetition of the amplitude fit.
The possible influence of additional resonant components is investigated by including an extra, statistically insignificant resonance—$\Lambda(1520)$—in the fit. 
The resulting variation in the fit results is assigned as a systematic uncertainty.
To account for the uncertainty associated with the polarization parameter, 500 pseudoexperiments are generated by Gaussian sampling of the fixed polarization values. 
The amplitude fit is repeated for each sample, and the standard deviation of the 500 results is taken as the corresponding systematic uncertainty.
Finally, the systematic uncertainty arising from differences between data and MC simulation is evaluated using correction factors that quantify efficiency discrepancies. 
These factors are defined as 
$\omega = \varepsilon_{\text{data}} / \varepsilon_{\text{MC}}$, 
where $\varepsilon_{\text{data}}$ and $\varepsilon_{\text{MC}}$ are the detection efficiencies obtained from control samples in data and MC simulation, respectively. 
The correction factors are applied to reweight the PHSP MC samples, and the full fit procedure is repeated to estimate the resulting systematic variation. The resulting systematic uncertainties of the amplitude analysis are listed in Table~\ref{tab:sys_err:pwa}.

\begin{table}[!htbp]
\begin{center}
    \caption{The relative systematic uncertainties in the branching fraction measurement.}
    \small
  \begin{tabular}{l c}
      \hline \hline
			Source            &  $\Lambda^{+}_{c}\to pK^{+}K^{-}~(\%)$   \\
			\hline
			Tracking      			                	&3.0		\\
			PID 					                	&3.0		\\
		$\Delta E$ requirement				        &0.2		\\
		  $M_{\rm BC}$ fit		                   	  &1.3		\\
			MC model				                 	&1.6		\\
		  $N_{\Lambda^{+}_{c}\bar{\Lambda}^{-}_{c}}$  &1.7		\\
			\hline
		    Total         			                 	&5.0		\\
      \hline \hline
    \end{tabular}
    \label{tab:sys_err:br}
\end{center}
\end{table}

The systematic uncertainties in the branching fraction measurement arise from the following sources: tracking efficiency, particle identification (PID), the total number of produced $\Lambda^{+}_{c}\bar{\Lambda}^{-}_{c}$ pairs ($N_{\Lambda^{+}_{c}\bar{\Lambda}^{-}_{c}}$), the $M_{\rm BC}$ fit, the $\Delta E$ selection requirement, and the MC signal model.
The uncertainties associated with the tracking and PID efficiencies of charged particles are estimated to be $1\%$ per track. 
These values are determined using control samples from the processes $e^+e^-\to\pi^+\pi^+\pi^-\pi^-$ and $e^+e^-\to K^+K^-\pi^+\pi^-$ collected at CM energies above $\sqrt{s}=4.0~\rm{GeV}$~\cite{BESIII:2022xne}. 
The uncertainty associated with the $M_{\rm BC}$ fit is evaluated to be $1.3\%$ by varying both the signal and background descriptions, where the signal shape is modified by convolving the signal MC shape with a double-Gaussian function and the background-shape parameters are fixed to those determined from background MC samples.
The uncertainty due to the $\Delta E$ selection requirement is estimated by varying the resolution of the $\Delta E$ distribution, resulting in an uncertainty of $0.2\%$.
For the MC signal model, the signal shape is derived from simulated samples generated using the amplitude model obtained in the nominal fit. 
To assess the model-related uncertainty, the amplitude parameters are randomly varied according to their nominal values and uncertainties in 100 iterations. 
For each iteration, the full fit procedure is repeated to re-evaluate the efficiency and branching fraction. 
The resulting branching fraction distribution is fitted with a Gaussian function, and the ratio of its standard deviation to the nominal branching fraction value ($1.6\%$) is assigned as the systematic uncertainty.
Finally, the uncertainty in the total number of $\Lambda^{+}_{c}\bar{\Lambda}^{-}_{c}$ pairs, $N_{\Lambda^{+}_{c}\bar{\Lambda}^{-}_{c}}$, is taken to be $1.7\%$, as quoted from Ref.~\cite{BESIII:2026qbp}.
The resulting systematic uncertainties in the branching fraction measurement are listed in Table~\ref{tab:sys_err:br}.

\section{Summary}
To summarize, based on $e^{+}e^{-}$ collision data corresponding to an integrated luminosity of 4.4~$\rm{fb}^{-1}$ collected with the BESIII detector at CM energies between 4600 and $4698~\rm{MeV}$, an amplitude analysis of the charmed baryon decay $\Lambda^{+}_{c}\to pK^{+}K^{-}$ has been performed. 
Using the amplitude model derived from this analysis, the relative branching fractions of the resonant components are measured for the first time. 

The amplitude solution shows that $\Lambda_c^+ \to pK^+K^-$ is not exhausted by a single $p\phi(1020)$ contribution. Instead, the baseline model contains $p\phi(1020)$, $pf_0(980)$, $\Lambda(1405)K^+$, and $\Lambda(1670)K^+$ components, with no evidence for a significant nonresonant contribution. Although the $p\phi(1020)$ component provides the largest contribution, the $pf_0(980)$ component is also significant.
Compared with the Cabibbo-favored mode $\Lambda_c^+ \to pK^-\pi^+$, the absence of visible resonant structures in the $pK^+$ system and the more limited phase space available in the $pK^-$ system lead to a simpler amplitude model and a clearer substructure pattern in the present singly Cabibbo-suppressed decay $\Lambda_c^+ \to pK^+K^-$.

Based on the amplitude-analysis results and the measurement of $\mathcal{B}(\Lambda^{+}_{c}\to pK^{+}K^{-}) = (1.08\pm0.05)\times10^{-3}$ reported by the PDG~\cite{ParticleDataGroup:2024cfk}, the absolute branching fractions of the individual resonant components are obtained, as summarized in Eq.~\ref{BF:result}.
The quoted uncertainties are, in order, statistical, systematic, and from the PDG reference value. 
Since the PDG provides averages with higher precision than our own result, it is adopted to reduce the overall uncertainty.
Furthermore, by incorporating the branching fraction of $\phi(1020)\to K^{+}K^{-}$, the branching fraction of $\Lambda^{+}_{c}\to p\phi(1020)$ is determined to be $(1.24\pm0.11\pm0.04\pm0.06)\times10^{-3}$, representing the most precise determination to date.
This value agrees well with theoretical predictions from the pole model~\cite{Zenczykowski:1993hw,Zenczykowski:1993jm} and the covariant confined quark model (CCQM)~\cite{Ivanov:1997ra}, and is consistent with previous experimental results and the PDG average.

\begin{widetext}
\begin{equation}
  \begin{split}
  \footnotesize
   \mathcal{B}(\Lambda^{+}_{c}\to\Lambda(1405)K^{+}, \Lambda(1405)\to pK^{-})&=(0.23\pm0.10\pm0.06\pm0.01)\times10^{-3}, \\
   \mathcal{B}(\Lambda^{+}_{c}\to\Lambda(1670)K^{+}, \Lambda(1670)\to pK^{-})&=(0.13\pm0.11\pm0.09\pm0.01)\times10^{-3}, \\
   \mathcal{B}(\Lambda^{+}_{c}\to pf_{0}(980), f_{0}(980)\to K^{+}K^{-})&=(0.43\pm0.17\pm0.16\pm0.02)\times10^{-3}, \\
   \mathcal{B}(\Lambda^{+}_{c}\to p\phi(1020), \phi(1020)\to K^{+}K^{-})&=(0.62\pm0.05\pm0.02\pm0.03)\times10^{-3}.
  \end{split}
  \label{BF:result}
\end{equation}
\end{widetext}

Benefiting from the amplitude analysis technique and the further increase in data statistics, the overall branching fraction of $\Lambda^{+}_{c}\to pK^{+}K^{-}$ is determined to be 
$(9.94\pm0.65\pm0.50)\times10^{-4}$. 
This result supersedes the previous BESIII measurement~\cite{BESIII:2016ozn} with precision improved by approximately a factor of 1.5. 
Our measurement is consistent with other experimental results, but deviates from the CLEO measurement, 
$(2.44\pm0.56\pm0.44)\times10^{-3}$~\cite{CLEO:1995moc}, by about $2.0\sigma$. 
Due to the limited statistical precision of the current data sample, both the branching-fraction measurements and the amplitude-model results remain statistically dominated and no decay-parameter measurements are reported in this work. Nevertheless, the obtained amplitude solution still provides a useful description of the $\Lambda_c^+ \to pK^+K^-$ decay dynamics and can serve as an important reference for future studies of local or global $CP$ asymmetries in charmed-baryon decays.

%\section*{Acknowledgement}
The BESIII Collaboration thanks the staff of BEPCII (https://cstr.cn/31109.02.BEPC) and the IHEP computing center for their strong support. This work is supported in part by National Key R\&D Program of China under Contracts Nos. 2023YFA1606000, 2023YFA1606704, 2023YFA1609400; National Natural Science Foundation of China (NSFC) under Contracts Nos. 11635009, 11935015, 11935016, 11935018, 12025502, 12035009, 12035013, 12061131003, 12105127, 12192260, 12192261, 12192262, 12192263, 12192264, 12192265, 12221005, 12225509, 12235017, 12361141819, 12422504; the Chinese Academy of Sciences (CAS) Large-Scale Scientific Facility Program; the Strategic Priority Research Program of Chinese Academy of Sciences under Contract No. XDA0480600; CAS under Contract No. YSBR-101; 100 Talents Program of CAS; Fundamental Research Funds for the Central Universities, Lanzhou University under Contracts Nos. lzujbky-2023-stlt01, lzujbky-2025-ytB01; The Institute of Nuclear and Particle Physics (INPAC) and Shanghai Key Laboratory for Particle Physics and Cosmology; ERC under Contract No. 758462; German Research Foundation DFG under Contract No. FOR5327; Istituto Nazionale di Fisica Nucleare, Italy; Knut and Alice Wallenberg Foundation under Contracts Nos. 2021.0174, 2021.0299; Ministry of Development of Turkey under Contract No. DPT2006K-120470; National Research Foundation of Korea under Contract No. NRF-2022R1A2C1092335; National Science and Technology fund of Mongolia; Polish National Science Centre under Contract No. 2024/53/B/ST2/00975; STFC (United Kingdom); Swedish Research Council under Contract No. 2019.04595; U. S. Department of Energy under Contract No. DE-FG02-05ER41374.

\end{document}